\documentclass{ieeeaccess}
\usepackage{makecell,balance}
\usepackage{caption}
\usepackage{url}
\usepackage{hyperref}
\hypersetup{
 colorlinks=True,
}
\usepackage{subfigure}
\usepackage{booktabs}
\usepackage{multirow}

\newcommand{\cmmnt}[1]{}
\usepackage[table,xcdraw, dvipsnames]{xcolor}
\usepackage{float}

\usepackage{cite}
\usepackage{amsmath,amssymb,amsfonts}
\usepackage{algorithmic}
\usepackage{graphicx}
\usepackage{textcomp}
\newcommand{\hl}[1]{#1}
\def\BibTeX{{\rm B\kern-.05em{\sc i\kern-.025em b}\kern-.08em
    T\kern-.1667em\lower.7ex\hbox{E}\kern-.125emX}}
\begin{document}
\newcommand{\etal}{\emph{et~al.}}
\history{Date of publication xxxx 00, 0000, date of current version xxxx 00, 0000.}
\doi{10.1109/ACCESS.2017.DOI}

\title{Heart Sound Classification Considering Additive Noise and Convolutional Distortion}


\author{\uppercase{Farhat~Binte~Azam$^1$}, \IEEEmembership{Student Member,~IEEE},
\uppercase{Md.~Istiaq~Ansari$^1$},\\
\uppercase{Ian~McLane$^2$}, \IEEEmembership{Member, IEEE} AND
\uppercase{Taufiq~Hasan$^1$} \IEEEmembership{Senior Member, IEEE}}
\address{$^1$mHealth Laboratory, Department of Biomedical Engineering, Bangladesh University of Engineering and Technology (BUET), Dhaka - 1205, Bangladesh.\\
$^2$Sonavi Labs Inc., Baltimore, MD 21230, USA.
}



\corresp{Corresponding author: Taufiq Hasan (e-mail: taufiq@bme.buet.ac.bd).}

\begin{abstract}
Cardiac auscultation is an essential point-of-care method used for the early diagnosis of heart diseases. Automatic analysis of heart sounds for abnormality detection is faced with the challenges of additive noise and sensor-dependent degradation. This paper aims to develop methods to address the cardiac abnormality detection problem when both types of distortions are present in the cardiac auscultation sound. We first mathematically analyze the effect of additive and convolutional noise on short-term filterbank-based features and a Convolutional Neural Network (CNN) layer. Based on the analysis, we propose a combination of linear and logarithmic spectrogram-image features. These 2D features are provided as input to a residual CNN network (ResNet) for heart sound abnormality detection. Experimental validation is performed on an open-access heart sound abnormality detection dataset involving noisy recordings obtained from multiple stethoscope sensors. The proposed method achieves significantly improved results compared to the conventional approaches, with an area under the ROC (receiver operating characteristics) curve (AUC) of 91.36\%, F-1 score of 84.09\%, and Macc (mean of sensitivity and specificity) of 85.08\%. We also show that the proposed method shows the best mean accuracy across different source domains \hl{including stethoscope and noise variability}, demonstrating its effectiveness in different recording conditions. The proposed combination of linear and logarithmic features along with the ResNet classifier effectively minimizes the impact of background noise and sensor variability for classifying phonocardiogram (PCG) signals. The proposed method paves the way towards developing computer-aided cardiac auscultation systems in noisy environments using low-cost stethoscopes.
\end{abstract}


\titlepgskip=-15pt

\maketitle

\PARstart{C}{ardiovascular} Diseases (CVDs) pose a significant burden on the public health sector, causing about 17.9 million deaths every year, which is 31\% of all deaths worldwide \cite{whofact}. According to the Center for Disease Control (CDC), in the US alone, one person dies of CVDs every 36 seconds \cite{HeartDis39:online}. With the shortage of trained physicians and scarcity of diagnostic equipment, the burden of CVDs is much higher in the low- and medium-income countries \cite{whofact}. Since early diagnosis is the key to reducing the burden of CVDs, low-cost screening tools such as digital stethoscopes with automatic murmur analysis algorithms are becoming more widely available \cite{leng2015electronic}. With the recent advancement in embedded systems, it is now possible to incorporate complex machine learning models within built-in microprocessors within the stethoscope \cite{west2019introducing, mclane2021design}. However, distortions introduced by the stethoscope sensor (e.g., diaphragm, amplifier) and background noise are of particular concern in designing computer-aided cardiac auscultation frameworks, especially in the underserved communities.

A considerable amount of research has been performed in the area of automatic heart sound analysis for the detection of CVDs. Publicly available datasets \cite{goldberger2000physiobank,schuller2017interspeech} have accelerated the progress of research in this area, in particular the 2016 Physionet/CinC Challenge \cite{clifford2016classification,liu2016open} dataset.
This dataset provides an archive of 4430 PCG recordings acquired using seven different stethoscopes/sensors and are annotated as either physiological (normal) or pathological (abnormal). Various approaches have already been attempted for heart sound abnormality detection using this dataset. 
A diverse set of front-end features have been examined including, time, frequency and statistical features \cite{homsi2017ensemble}, Mel-frequency Cepstral Coefficients (MFCC) \cite{rubin2016classifying,bobillo2016tensor, bozkurt2018study,noman2019short}, and Continuous Wavelet Transform (CWT) \cite{ren2018learning, kay2017dropconnected} features. 
\hl{Time-frequency based features including MFCC and Mel-Spectrograms are most commonly being used in recent studies.}
A wide array of classifiers have been attempted including the k-Nearest Neighbor (k-NN) \cite{bobillo2016tensor}, Support Vector Machine (SVM) \cite{whitaker2017combining}, Random Forest \cite{homsi2017ensemble}, Multilayer Perceptron (MLP) \cite{kay2017dropconnected}, \cite{zabihi2016heart}, deep learning approaches with 1D \& 2D CNNs \cite{potes2016ensemble, humayun2020towards, maknickas2017recognition, humayun2018learning, humayun2018ensemble}, and Recurrent Neural Network (RNN) \cite{yang2016classification}. While \cite{humayun2020towards} addressed the issue of domain variability for heart sound classification, \hl{the issue of additive noise was not addressed. A summary of recent research works relevant to the proposed method are provided in Table \ref{tab:related_work}. Note that different authors used different train-test splits using the PhysioNet data, thus, the results are not comparable in most cases.}

\begin{table*}[]
    \centering
    \caption{Summary of recent research works on heart sound classification using 2016 PhysioNet/CinC Database \cite{liu2016open}}
    \label{tab:related_work}
    \begin{tabular}{c|c|p{1.2cm}|p{1.7cm}|c|c|c|c|p{4cm}}
    \toprule
        Author & Year & Features & Classifier & Sen. & Spe. & MAcc. & Acc. & Remarks \\
        \hline 
        Potes~\etal \cite{potes2016ensemble} & 2016 & Raw waveform & Branched 1D-CNN& 0.88 & 0.82 & 0.85 & - & Top scoring system in \cite{clifford2016classification}. In-house test performance reported.\\\hline
        Rubin~\etal \cite{rubin2016classifying} & 2016 & MFCC & 2D-CNN & 
        0.765 & 0.931 & 0.848 & - & Official test set results are reported (not publicly available). \\\hline
        Bozkurt~\etal \cite{bozkurt2018study} & 2018 & MFCC & 2D-CNN & 0.845 & 0.785 & - & 0.815 & Noise and/or sensor variability effects not considered.\\\hline
        Ren~\etal \cite{ren2018learning} & 2018 & Scalogram & Pre-trained 2D-CNN & 0.246 & 0.878 & 0.562 & - & Noise and/or sensor variability effects not considered.\\\hline
        Noman~\etal \cite{noman2019short} & 2019 & MFCC & 1D-CNN, 2D-CNN& 0.8994& 0.8635 & 0.8815 & 0.8922 & Custom data split used. Domain variability or noise not considered.\\\hline
        Humayun~\etal \cite{humayun2020towards} & 2020 & Raw waveform & Learnable filter \& 1D-CNN & 0.8695$^*$  & 0.7602$^*$  & 0.8149$^*$  & - & Addressed domain variability. Did not consider additive noise.\\\hline
        Deng~\etal \cite{deng2020heart} & 2020 & MFCC & Recurrent 2D-CNN &
        0.9866 & 0.9801 & - & 0.9834 & 10-fold cross-validation used which shows overoptimistic results.\\\hline
        \bottomrule
    \end{tabular}
    \vspace{2mm}
    \\$^*$This study uses a balanced test data across domains and pathology class using the 2016 PhysioNet validation set. This allows for better evaluation of the performance across different data subsets. However, the results are not comparable with other studies using different data-splits or a 10-fold cross validation.
\end{table*}

\begin{figure}[t]
\includegraphics[width=\linewidth]{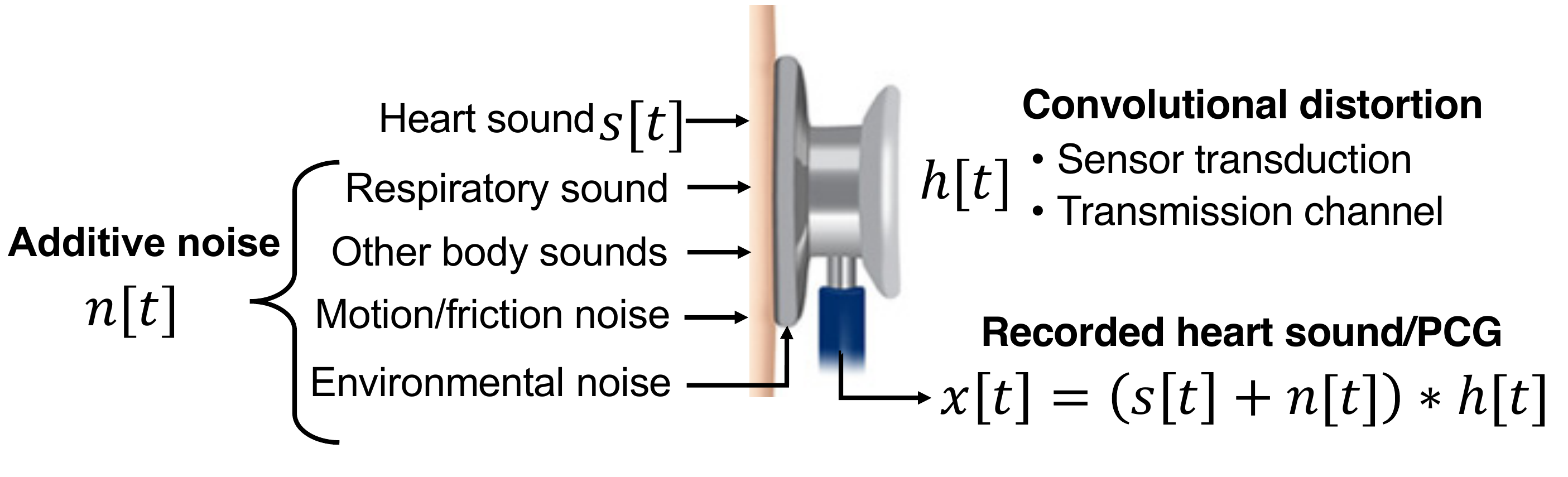}
\centering
\caption{Schematic model showing the sources of additive noise and convolutional distortion during cardiac auscultation.}
\label{signal}
\vspace{-3mm}
\end{figure}

\hl{A major limitation of the previous approaches is that while only sensor variability was addressed in \cite{humayun2020towards}, none of the works have considered the simultaneous presence of channel/sensor and additive noise distortion for this task. Sensor/channel variability affects the signal through time-convolution, whereas additive noise affects the signal linearly. Thus it is very difficult to disentangle the two components of distortion. Traditional signal processing methods have addressed mainly the channel distortion issue through the log-filterbank analysis, where the convolutional distortion becomes additive \cite{oppenheim2004frequency}. This also explains the effectiveness of MFCC features for the heart sound classification task. Previous work on addressing simultaneous presence of additive and convolutional noise has been mainly focused on speech processing applications  \cite{acero2000hmm,stouten2004joint,gong2005method}. However, this issue is yet to be addressed for heart sound analysis in the literature.} 

In the case of heart sounds, the \hl{major sources of signal degradation are: (i)} the variability due to stethoscopes/sensor \cite{humayun2020towards} affecting the PCG as a convolutional degradation (channel distortion), and \hl{(ii)} the effect of additive noise (e.g., from the environment or the patient). These effects are depicted in Fig. \ref{signal}. \hl{In particular, noise is a major problem in automatic assessment of cardiac diseases using stethoscopes, especially in low-resource hospitals.} In \cite{humayun2020towards}, it was shown that the state-of-the-art algorithms for heart sound classification are not capable of providing a consistent performance when different stethoscopes are encountered in the test data. While the overall performances were impressive, the algorithms easily overfit to the data recorded using the most prevalent stethoscope. \hl{This is the reason why 10-fold cross-validation \cite{deng2020heart}, frequently performed on this dataset, provides overoptimal performance as the test sets cannot provide sufficient variability due to the severe domain imbalance present in the Physionet data (Fig. \ref{fig:data}).}
In \cite{humayun2020towards}, a mini-batch balancing method was proposed to address the performance issue due to domain variability, and consistent improvement of performance across different domains was found \hl{on a domain and class balanced test set.} \hl{However, this dataset also contains additive noise due to human speech, environmental sounds, stethoscope motion, breathing, intestinal activity, etc. \cite{liu2016open}, which was not considered in \cite{humayun2020towards}.}

\hl{This paper aims to address the simultaneous presence of additive noise and convolutional distortion for the heart sound classification problem. In the case of cardiac auscultation, additive noise mainly includes respiratory sounds and environmental noise. The convolutional distortion represents the transduction effect of the stethoscope sensor. The main contributions of this paper are:
\begin{itemize}
    \item We introduce a model of the acoustic environment during auscultation, considering the additive and convolutional distortions present in the system.
    \item Perform analysis of additive and convolutional distortion on heart sounds in the linear and logarithmic filterbank domain and show that both distortions can be converted into an additive distortion.
    \item Analyze the effect of additive uncorrelated random noise in the first layer of a conventional CNN model.
    \item Proposed a simple feature fusion technique combining linear and logarithmic filterbank features to simultaneously address additive noise and convolutional distortion.
    \item Experimentally validate that the proposed feature fusion scheme can effectively improve heart sound classification performance compared to existing methods using a deep residual network.
\end{itemize}
}


The remainder of this paper is organized as follows. In Section \ref{motive}-\ref{proposed} we elaborate on our motivation and implementation details of our proposed method. Section \ref{expres} describes the datasets used, different experiments performed, analysis of the results obtained by the proposed method and performance comparisons. Finally, in Section \ref{disc} and \ref{conc}, we discuss the limitations of our method, future directions and summarize our findings.

\begin{figure}[t]
\includegraphics[width=\linewidth]{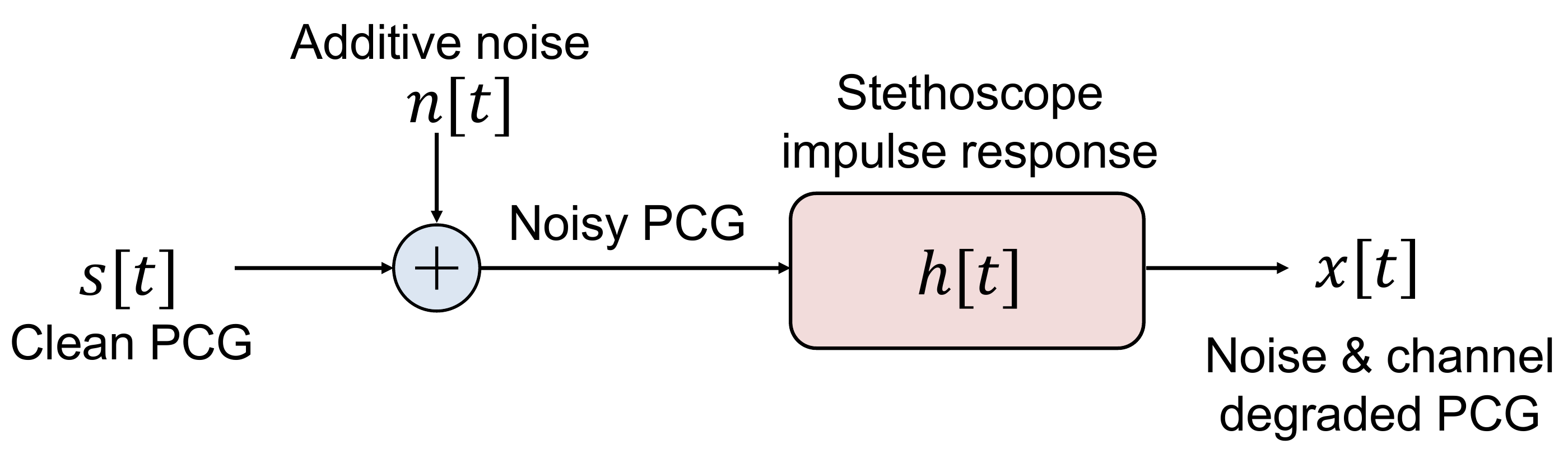}
\centering
\caption{Parametric model of the PCG signal acquisition process showing additive noise and convolutional distortion.}
\label{signal1}
\vspace{-3mm}
\end{figure}

\section{{Background}} \label{motive}
\subsection{Additive and convolutional distortion}
Inspired by \cite{acero2000hmm}, we consider a parametric model of the stethoscope as shown in Fig. \ref{signal1}. This assumes that the original PCG signal and an uncorrelated noise component are first added before being processed through a time-convolution operation due to the stethoscope's impulse response. For a given short-time segment of length $N$, we assume that the clean PCG, noisy PCG, additive noise and stethoscope impulse response are given by, $s[t]$, $x[t]$, $n[t]$ and $h[t]$, 
respectively. According to the parametric model of Fig. \ref{signal1},
\begin{equation} \label{eq1}
    x[t] = \Big(s[t] + n[t] \Big) * h[t].
\end{equation}
Note that we assume the stethoscope impulse response is independent of the short-time segment as it is a time-invariant property of the device. We also assume that the impulse response $h[t]$ has a length less than $N$. Expressing (\ref{eq1}) in the Discrete Fourier Transform (DFT) domain, in each segment,  using the convolution property of DFT, we obtain 
\begin{equation} \label{eq2}
    X[k] = \Big(S[k] + N[k]\Big) H[k].
\end{equation}
In this step, we assume a $K$-point DFT with $k = 0,1\cdots,K$ representing the frequency indices, whereas $X[\cdot]$, $S[\cdot]$, $N[\cdot]$ and $H[\cdot]$ represent the DFT coefficients of the corresponding time-domain signals. Note that the convolutional distortion $H[k]$ due to the stethoscope transfer function has now become multiplicative.
Next, we estimate the power spectrum of this short-time PCG segment by squaring the absolute value of the DFT coefficients and obtain
\begin{equation} \label{eq3}
    \begin{split}
        \left|X[k]\right|^2 & = \Big(|S[k]|^2 + |N[k]|^2\Big)|H[k]|^2 \\
        & + 2\mbox{Re}\{S[k] H[k] N[k]^* H[k]^*\}.
    \end{split}
\end{equation}

\subsection{Linear filterbank energy analysis}
Audio signals can be conveniently analyzed in different frequency bands by using a filterbank. Let us assume a pre-defined filterbank of $M$ filters and denote the $i$-th filter energy coefficient of the noisy PCG signal as $X[i]$. We assume that the filterbank-energy term $X[i]$ is related to $X[k]$ by the expression
\begin{equation}
    |X[i]|^2 = \mathbb{E}^i_k \Big(|X[k]|^2\Big).
\end{equation}
Here, $\mathbb{E}^i_k$ indicates that the expected value operation is performed over the $k$ frequency indices that belong to the $i$-th filter. Substituting the values from (\ref{eq3}) we obtain,
\begin{eqnarray}
    |X[i]|^2 &=& \mathbb{E}^i_k \Big(|S[k]|^2|H[k]|^2 + |N[k]|^2|H[k]|^2 \nonumber\\
    &&+ 2\mbox{Re}\{S[k] H[k] N[k]^* H[k]^*\}\Big).\label{eq5}
\end{eqnarray}
Statistically, heart sound $s[t]$ and additive noise $n[t]$ are independent and uncorrelated, and therefore the expected value of the last term of (\ref{eq5}) can be approximated to be zero. According to this assumption and using the definition of filterbank energy, we simplify (\ref{eq5}) to obtain a similar expression derived in \cite{acero2000hmm} given by
\begin{equation}\label{eq_fbank_energy}
    |X[i]|^2 = (|S[i]|^2 + |N[i]|^2) |H[i]|^2.
\end{equation}
Here, $S[i]$, $N[i]$ and $H[i]$ represent the filterbank-energy values obtained from the corresponding DFT coefficients. Thus, according to (\ref{eq_fbank_energy}), the background noise component $N[i]$ is additive to the signal component $S[i]$ whereas the channel distortion component $H[i]$ is multiplicative in the filterbank energy domain. 

\subsection{Log-filterbank energy analysis}
In filterbank analysis, logarithm operation is usually performed on the energy coefficients to increase the dynamic range and transform multiplicative operation into addition. Taking natural logarithm over (\ref{eq_fbank_energy}) we obtain
\begin{equation} \label{eq_logfbank_energy}
        \log |X[i]|^2 = \log|H[i]|^2 + \log\Big( |S[i]|^2+|N[i]|^2 \Big). 
\end{equation}
As expected, (\ref{eq_logfbank_energy}) reveals that in the log-filterbank domain, the channel or transducer/sensor distortion becomes additive while the heart sound signal and background noise becomes non-linear functions of the output log-energy coefficient. This was one of the motivations of Cepstral analysis to separate the channel effect from the signal \cite{oppenheim2004frequency,deller2000discrete}. We may write (\ref{eq_logfbank_energy}) to express the log-energy vectors for each frame using matrix notation as
\begin{equation}\label{eq_logfbank_vec}
     \mathbf{E}_X = \mathbf{E}_H + \mathbf{E}_{S+N}, 
\end{equation}
where 
\begin{eqnarray}
    \mathbf{E}_X &=& \Big[\log |X[1]|^2, \log |X[2]|^2 \cdots \log |X[M]|^2\Big]^T,\nonumber\\
    \mathbf{E}_X &=& \Big[\log |H[1]|^2, \log |H[2]|^2 \cdots \log |H[M]|^2\Big]^T, \mbox{ and}\nonumber\\
    \mathbf{E}_{S+N} &=& \left[ 
        \begin{array}{c}
        \log(|S[1]|^2+|N[1]|^2)\\
        \log(|S[2]|^2+|N[2]|^2)\\
        \vdots\\
        \log(|S[M]|^2+|N[M]|^2)\\
        \end{array} 
        \right].\nonumber
\end{eqnarray}
In (\ref{eq_logfbank_vec}), each vector consists of $M$ elements obtained from the individual front-end filters. To summarize our analysis on filterbank and log-filterbank energy analysis, when both additive and convolutional distortion is present, it is not possible to decompose them into additive terms in the feature domain. However, noise component as in (\ref{eq_fbank_energy}) or the convolutional distortion component as in (\ref{eq_logfbank_energy}) can become additive separately. A static additive distortion component from features can be easily reduced by mean normalized which was the original motivation for the traditional Cepstral mean normalization (CMN) \cite{liu1993efficient} used for robust speech recognition.

\subsection{cepstral feature domain analysis}
MFCC features are frequently utilized in speech and audio processing. After the extraction of log-filterbank features, an additional step of Discrete Cosine Transform (DCT) is applied to obtain the well-known MFCC coefficients \cite{oppenheim2004frequency}. The DCT step is performed to reduce the correlation among the different filterbank energy coefficients and compress the overall energy of the features to the leading few components. Therefore, after this step, the higher-order DCT coefficients are usually eliminated, and thus, the feature dimensionality is reduced. Since DCT is a linear operation, it follows from (\ref{eq_logfbank_vec}) that the convolutional distortion component will still be an additive component after this operation. Thus we have
\begin{equation}\label{eq_MFCC_additive}
    \mbox{DCT}\left(\mathbf{E}_X\right) = \mbox{DCT}\left(\mathbf{E}_H\right) + \mbox{DCT}\left(\mathbf{E}_{S+N}\right)
\end{equation}
Similarly, we can easily show that by removing the higher order DCT coefficients, (\ref{eq_MFCC_additive}) still holds for the remaining coefficients. However, the PCG signal and additive noise is still entangled within the component $\mathbf{E}_{S+N}$ where their relationship is non-linear. 

\subsection{Additive distortion in a CNN model}
Although CNNs are primarily used for image classification, in many 1D signal processing applications, 2D spectrogram-images are used as input features to CNN models for classification \cite{hyder2017acoustic}.
In addition, previous methods have utilized additive noise in training for data augmentation that generally improves the robustness of the models \cite{tran2017bayesian, mikolajczyk2018data}. In this sub-section, we perform a mathematical analysis of the CNN input layer when an additive noise component is present in the input feature data.

In the $0^{th}$ (first) layer of a CNN model, for a 2D input feature matrix $\mathbf{F}_{cn}$ and the $j$-th 2D kernel $\mathbf{K}_j$, the output $j$-th feature map before the nonlinear activation layer is given by
\begin{equation} \label{eq7}
        \mathbf{G}_j = \mathbf{F}_{cn} \otimes \mathbf{K}_j.
\end{equation}
Here, $\otimes$ represents the correlation operation performed in a CNN layer. Eq. (\ref{eq7}) can be expressed for each individual matrix elements of row $m$ and column $n$ as,
\begin{eqnarray} \label{eq7.1}
        G_j[m,n] &=& F_{cn}\otimes K [m,n]\nonumber\\
        &=& \sum_{\alpha=-l}^{l} \sum_{\beta=-l}^{l} F_{cn}[m+\alpha, n+\beta] K_j[\alpha, \beta].
\end{eqnarray}
Here, we assume a kernel size of $(2l+1)\times(2l+1)$, and $\alpha$ and $\beta$ are dummy variables. When spectrogram images from heart sounds are considered, $m$ represent the short-time frame index whereas $n$ represent the frequency-band index. 

Let us now assume that the input feature matrix $\mathbf{F}_{cn}$ can be expressed as
\begin{equation}
    \mathbf{F}_{cn} = \mathbf{F}_{c} + \mathbf{F}_{n} 
\end{equation}
where $\mathbf{F}_{c}$ is dependent on the class $c \in \mathcal{C}$ where $\mathcal{C}$ defines the set of disease classes, and $\mathbf{F}_{n}$ is an additive noise component independent and uncorrelated to $\mathbf{F}_{c}$ (and thus independent of $c$). The source of $\mathbf{F}_{n}$ may environmental noise or sensor degradation. Considering the presence of this additive distortion, we obtain from (\ref{eq7})
\begin{eqnarray}\label{eq8}
        \mathbf{G}_j &=& (\mathbf{F}_{c} + \mathbf{F}_{n}) \otimes \mathbf{K}_j\nonumber\\
        &=& \mathbf{F}_{c} \otimes \mathbf{K}_j + \mathbf{F}_{n} \otimes \mathbf{K}_j.
\end{eqnarray}
Assuming that the model has converged after being trained on a sufficiently large amount of data, we may compute the expected value of the feature map as
\begin{eqnarray}\label{eq8.1}
        \mathbb{E}\{\mathbf{G}_j\} & = \mathbb{E}\{\mathbf{F}_{c} \otimes \mathbf{K}_j\} + \mathbb{E}\{\mathbf{F}_{n} \otimes \mathbf{K}_j\}.
\end{eqnarray}
At this stage, we will make a few simplifying assumptions. Since CNN models are trained using a loss-function that minimizes the classification error, it can be expected when trained on sufficient training data, the model parameters will be dependent on the disease classes $c$. In contrast, the additive component $\mathbf{F}_{n}$ by definition is independent and uncorrelated of $\mathbf{F}_{c}$ and thus the disease class $c$. Therefore, we may assume that after the training is converged, $\mathbb{E}\{\mathbf{F}_{n} \otimes \mathbf{K}_j\} \approx 0$ 
and (\ref{eq8}) can be written as
\begin{eqnarray}\label{eq9}
        \mathbb{E}\{\mathbf{G}_j\} \approx \mathbb{E}\{\mathbf{F}_{c} \otimes \mathbf{K}_j\}.
\end{eqnarray}
Thus, we may conclude that in case of an additive uncorrelated noise present in the data, CNN model parameters can be assumed to be independent of the noise when sufficient training data is present. In contrast, for multiplicative noise in the feature matrix, it is obvious from (\ref{eq7.1}) that a decomposition similar to (\ref{eq8.1}) will not be possible and the learning of the network will be directly affected by the distortion. 

\subsection{Fusion of linear and logarithmic features}\label{motivation_fuse}
It is clear that linear filterbank energy features are linearly affected by the additive noise and non-linearly affected by channel/sensor noise, as shown in (\ref{eq_fbank_energy}). In contrast, logarithmic filterbank energy features are linearly affected by channel/sensor noise and non-linearly affected by additive noise, as demonstrated in (\ref{eq_logfbank_energy}). \hl{Thus, in this respect, the linear and logarithmic filterbank energy features are complimentary. However, if we concatenate the two types of features, both additive noise and convolutional distortion can be separated as an additive component in the fused feature space.}
\hl{As shown in (\ref{eq9}), a CNN model is less affected by additive distortion during classification, assuming that the noise is independent and uncorrelated with the signal component. However, in the proposed fused feature space, both noise and channel distortion components become additive. Therefore, we hypothesize that the fused feature set will provide} improved performance in the presence of both additive and convolutional/channel noise. \hl{The following sections of this paper will focus on designing the relevant features and validating this hypothesis through experimental evaluation.}

\section {{Proposed Method}} \label{proposed}
In this section, we discuss the development of the proposed feature extraction method and the CNN architecture for effective classification of heart sounds, considering both additive and convolutional distortions.

\subsection{Pre-processing and segmentation}
The heart-sound segments are first re-sampled to $1000$Hz, followed by band-pass filtering between $25$-$400$Hz, and cardiac cycle segmentation according to \cite{springer2015logistic}. In the next step, we ensure that each heart sound training segment consists of a single cardiac cycle of a fixed duration of $2.5$s. Zero-padding is applied if the cardiac cycle is shorter than $2.5$s. 
\subsection{Feature extraction framework}
\begin{figure}[b]
\centering 
\includegraphics[width=\linewidth,trim={0 0 0 0},clip]{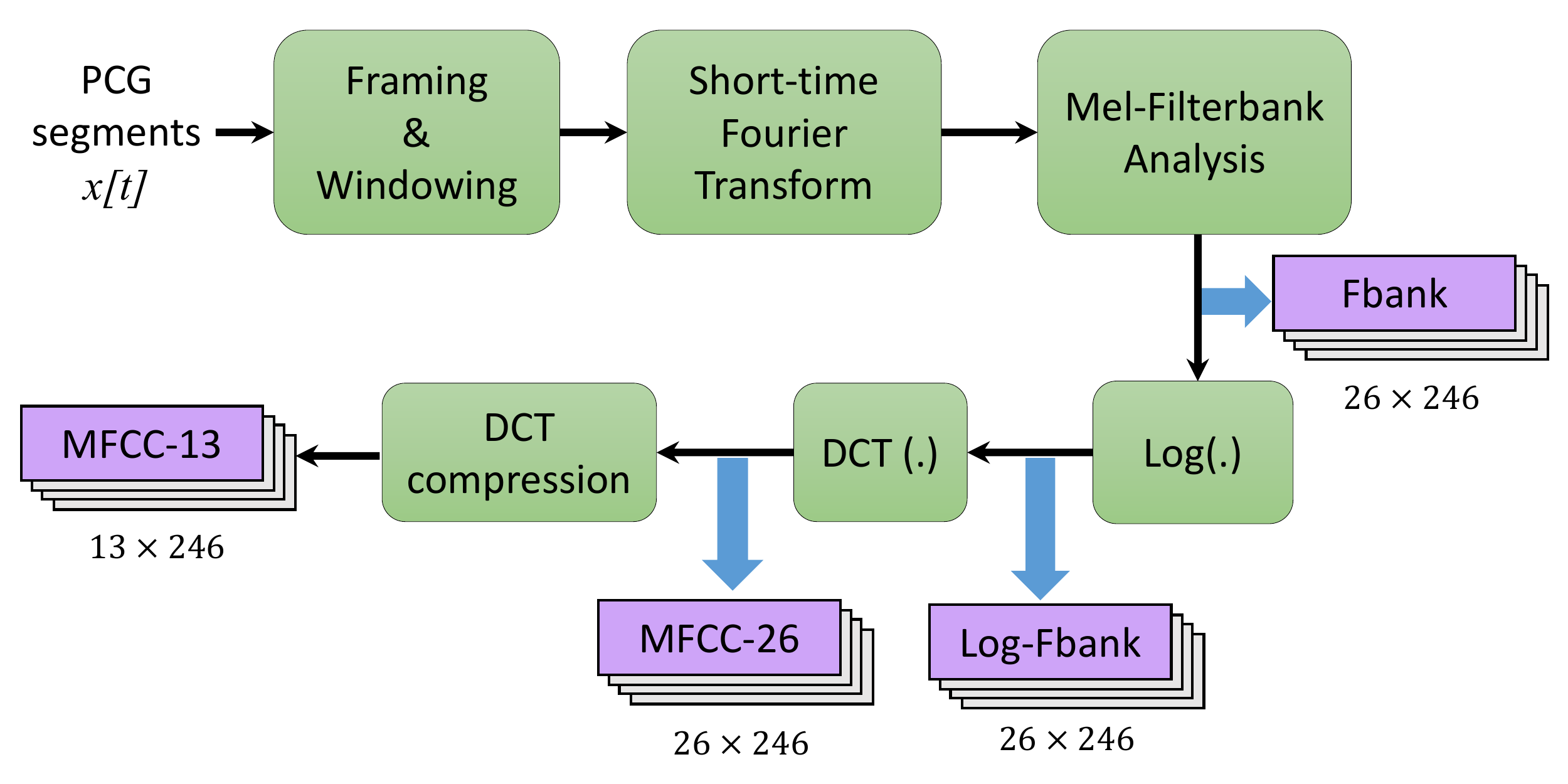}
\caption{A flow diagram of the acoustic features extraction framework. Different features are extracted from different stages as depicted.}
\label{fig:MFCC_flow}
\end{figure}

In order to validate our hypothesis, we needed to extract four different varieties of filter-bank based acoustic features from the PCG signal segments. These are based on equations (\ref{eq_fbank_energy}), (\ref{eq_logfbank_energy}) and (\ref{eq_MFCC_additive}). The features can be effectively extracted using a unified framework outlined in Fig. \ref{fig:MFCC_flow}. The extraction procedure is detailed below.

First, the PCG signal segments are divided into short-time frames of length $0.5$s with a hop size $0.01$s between successive frames. Filterbank analysis is performed on the $0.5$s segments using $26$ mel-scaled filters. The features extracted from these segments are summarized in Table \ref{table_features}.
\begin{table}[bht]
\centering
\caption{Description of acoustic features}
\label{table_features}
\resizebox{\linewidth}{!}{
\begin{tabular}{c|p{3.5cm}|c}
\toprule
\bf{Name} & \bf{Feature description} & \bf{Dimension}\\
\hline
Fbank & Filterbank energy features as shown in eq. (\ref{eq_fbank_energy}) & 26\\ 
Log-Fbank & Log-filterbank energy features as shown in eq. (\ref{eq_logfbank_energy}) & 26\\  
MFCC-26 & MFCC retaining all coefficients after DCT compression & 26\\
MFCC-13 & MFCC retaining 13 coefficients after DCT compression& 13\\
\bottomrule
\end{tabular}}
\end{table}
To validate our hypothesis, it is essential to perform classification experiments on all these features for the following reasons. According to (\ref{eq_fbank_energy}), the \emph{Fbank} features should be able perform better in case of additive noise. In the case of \emph{Log-Fbank}, we can expect better results in case of convolutional distortion as per our analysis in (\ref{eq_logfbank_energy}). Since CNN models are able to model the correlation among the input features, it is unclear if the DCT step will improve the performance in our case. For this reason, we retain both \emph{MFCC-13} and \emph{MFCC-26} variants of the traditional MFCC features to observe the effect of DCT in the classification performance.
\subsection{Preparing 2D feature matrices}
\begin{figure}
\centering 
\includegraphics[width=\linewidth] {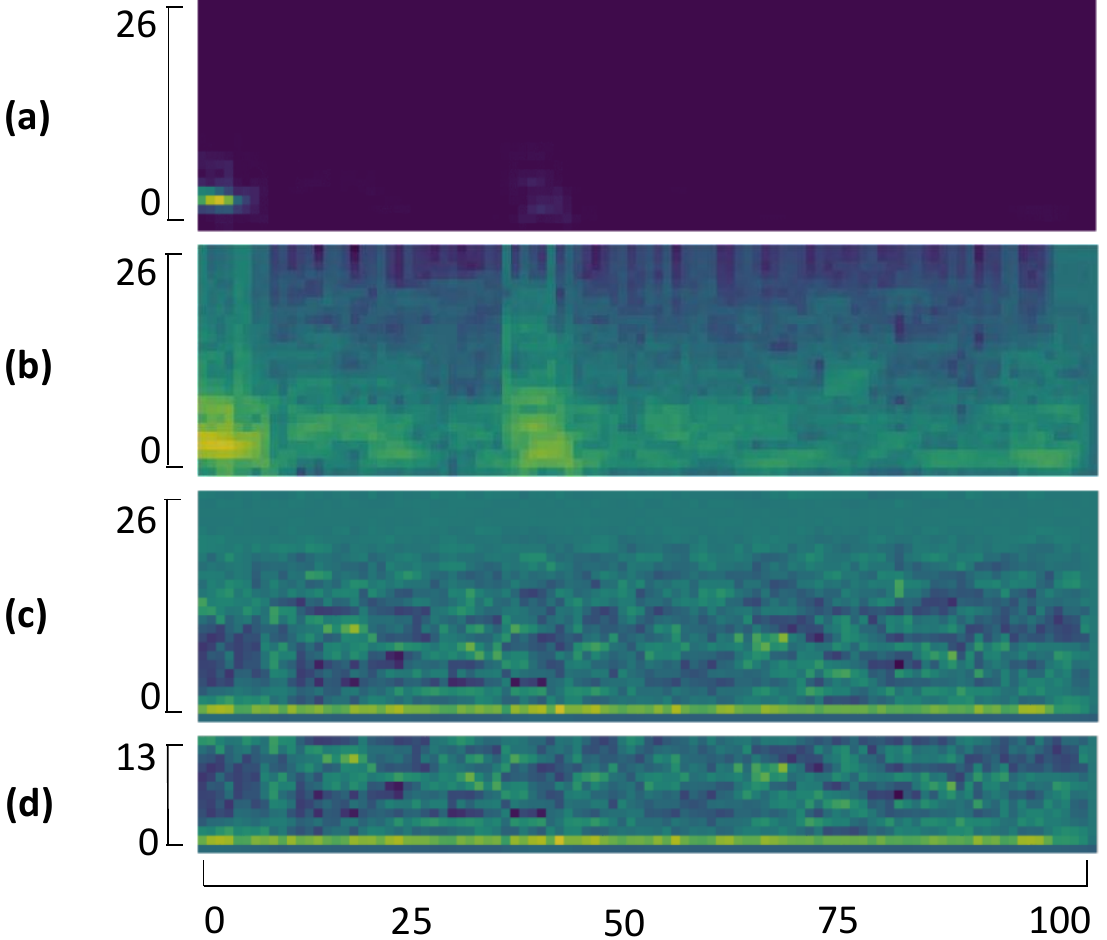}
\caption{2D visualization of the acoustic features as input to the proposed CNN model. (a) Filterbank energy (Fbank), (b) Log-Filterbank energy (log-Fbank), (c) Full-dimension MFCC (MFCC-26), (d) Dimension reduced MFCC (MFCC-13).}
\label{fig:single_features}
\vspace{3mm}
\end{figure}

From each $2.5$s segment, the extracted feature (column) vectors are stacked horizontally to form a 2D feature matrix (input) for classification using our CNN model. Each of these input feature matrices has a dimension of $d\times 246$, where $d$ represents the acoustic feature dimension. The four features as described in Table \ref{table_features} are visualized in Fig. \ref{fig:single_features}. Before feeding into the CNN model, features from each cardiac cycle are normalized by subtracting the mean of the full cycle and dividing by their standard deviation (STD).

\begin{figure}[tbh]
\centering 
\includegraphics[width=0.9\linewidth]{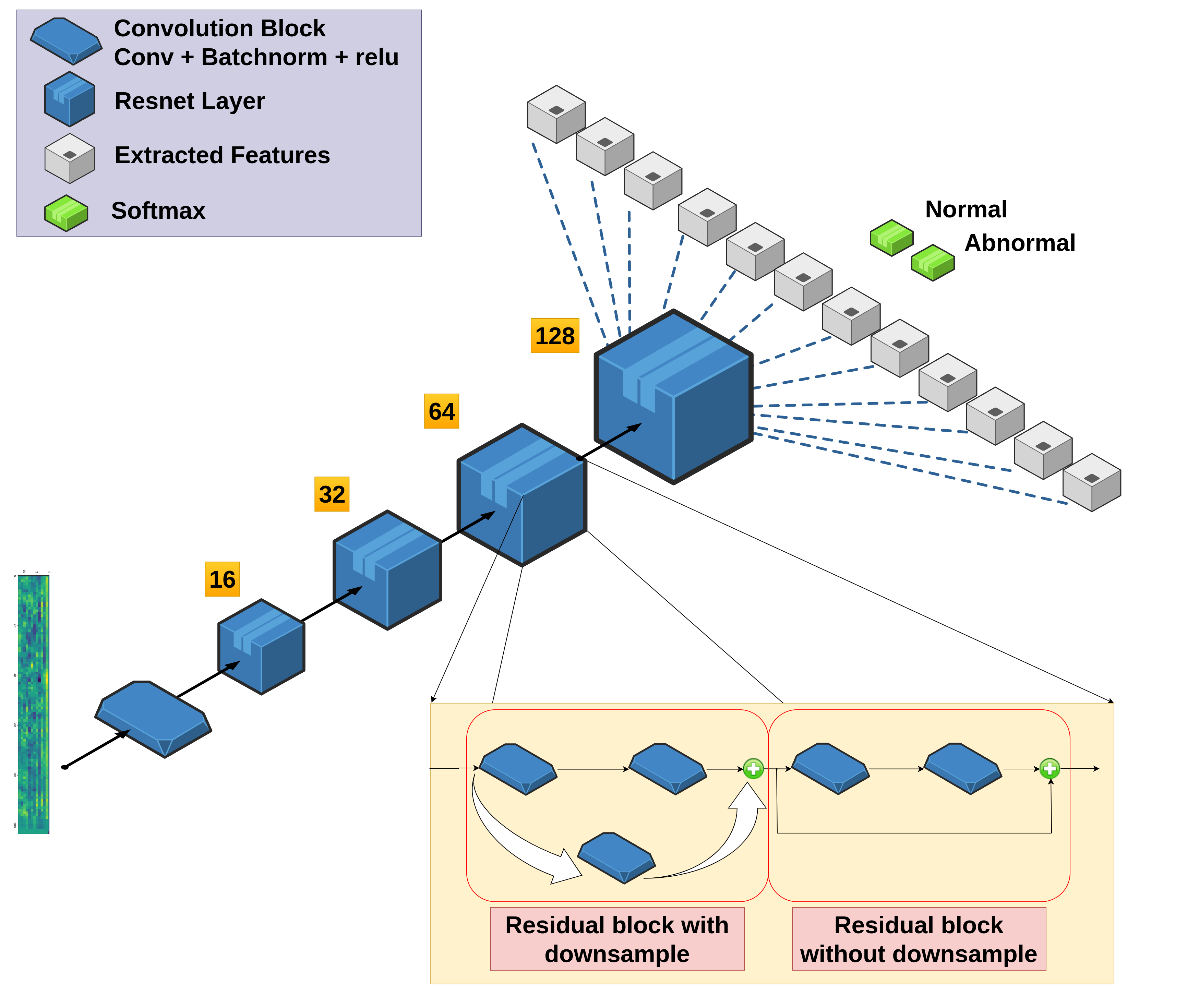}
\caption{Residual model architecture used for heart sound abnormality detection.}
\label{fig:model}
\vspace{-4mm}
\end{figure}

\subsection{Model architecture}
\hl{A residual neural network (ResNet) architecture based on \cite{he2016deep} is used as the classification model. The model is modified to fit the proposed input features}. The model has four resnet layers, each containing two residual blocks, the first block is with down-sampling, and the latter is with direct skip connection. Fig. \ref{fig:model} shows the model architecture used. Each residual block generally consists of two convolution blocks, which also perform the down-sampling operation. Each convolution block has one convolutional layer of kernel size $3$, followed by a 2D batch normalization layer and Rectified Linear Unit (ReLU) activation functions. Weights of each convolutional layer have are initialized using the Xavier method \cite{glorot2010understanding}. The number of filters of these four resnet layers is $16$, $32$, $64$, $128$. The input feature matrix goes through a convolutional block with a max-pooling of $2$ before the resnet layer. Extracted features from the fourth resnet layer are then flattened, and two neurons are added with softmax activation for binary classification (normal and abnormal PCG). \hl{The code is available online at \url{https://github.com/mHealthBuet/CepsNET}}.

\section{Experiments and results} \label{expres}
\subsection{Dataset}
We use the 2016 PhysioNet/CinC Challenge Database \cite{liu2016open} for our experiments. It is an archive of PCG recordings from both clinical and non-clinical environments. It consists of a total of 3,157 recordings from 764 patients and contains a total of 84,425 cardiac cycles ranging from heart rates 35 to 159 beats-per-minute (bpm). The PCG recordings have been collected from seven different research groups denoted as \{a-g,i\}. A brief summary of this dataset is shown in Table \ref{data}. Among seven categories, six \{a-f\} sets are distributed in training data, which is publicly available. The recordings generally have a duration between 5--120 seconds. 
Metadata of recording quality, annotations for the onset of S1, S2, systole, diastole are also provided. 

\hl{As shown in Table \ref{data}, this dataset is unbalanced both in terms of pathology-class (normal vs. abnormal) and the data subsets/domains (\{a-f\}). 
We use 2,856 (90.47\%) for training and the remaining 301 (9.53\%) for test. This test set is based on the original Physionet challenge validation set which is domain and class balanced \cite{humayun2020towards}. The train-test split is also patient independent by design.}
The class imbalance, variability of stethoscope sensors, and noisy environment are all challenging issues to be addressed in this dataset, making it suitable for experimental validation of the proposed approach. 
\renewcommand{\arraystretch}{1.5}
\begin{table}[h]
    \centering
    \caption{Data distribution of PhysioNet/CinC Challenge Database.}
    \resizebox{\linewidth}{!}{
        \begin{tabular}{@{}ccccc@{}}
            \toprule 
            \textbf{Subset} & 
            \textbf{\makecell{Total \\ Subject}} & 
            \textbf{\makecell{Normal \\ recordings}} & 
            \textbf{\makecell{Abnormal \\ recordings}} & 
            \textbf{Used device} \\\midrule
            a & 121 & 117 & 292 & Welch Allyn Meditron \\ 
            b & 106 & 385 & 104 & 3M Littmann E4000 \\
            c & 31 & 7 & 24 & AUDIOSCOPE \\
            d & 38 & 27 & 28 & Infral Corp. Prototype \\
            e (Norm.)  & 174 & 1867 & 0 & MLT201/Piezo \\
            e (Abn.) & 335\footnotemark & 0 & 151 & 3M Littmann \\
            f & 112 & 80 & 34 & JABES \\\bottomrule
        \end{tabular}}
    \label{data}
\end{table}
\footnotetext{We only used 151 usable abnormal recordings from 'e' subset and the rest are removed due to data errors \cite{humayun2020towards}.}

\subsection{Training regime}
\begin{figure}
\centering 
\includegraphics[width=\linewidth]{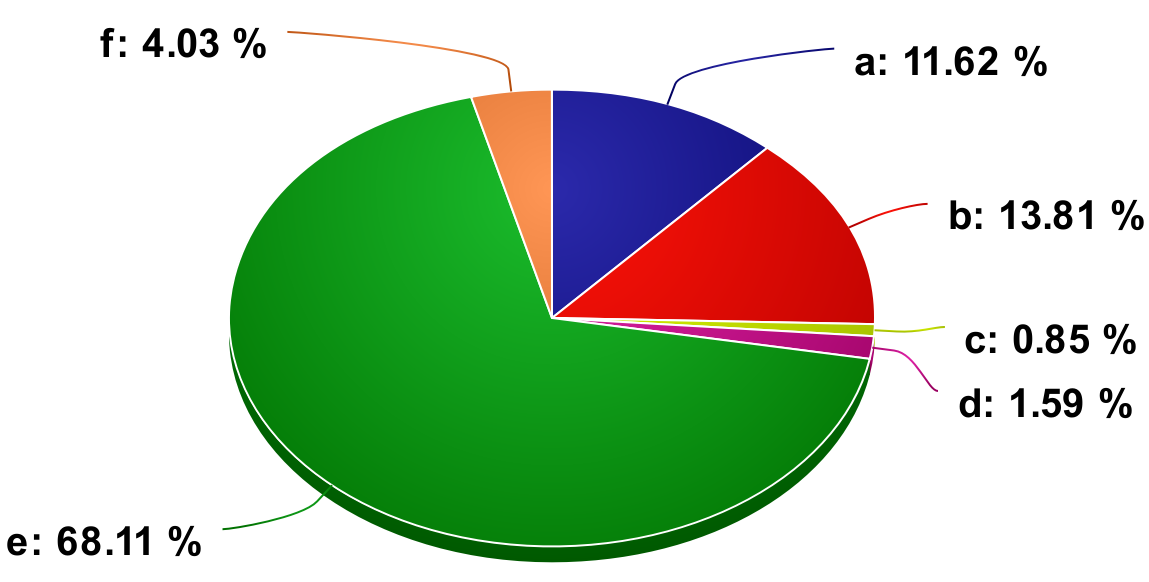}
\caption{Distribution of sensor-dependent subsets of the PhysioNet heart sound data.}
\label{fig:data}
\end{figure}

\hl{In the training-set, the e-subset has the highest number of recordings (about 68\%) as can be seen in Fig. \ref{fig:data}. Thus }we use the domain balance training (DBT) method that prepares mini-batches of data that are balanced for both class and domain \cite{humayun2020towards}. This method was found to be effective in case of domain variability and data imbalance.
We also apply shifting on both right and left directions to avoid the effect of segmentation errors.

To examine the effect of the input feature variation on classification performance, we perform our experiments in two steps. First, we use 2D matrices extracted from the individual features described in Table \ref{table_features} as input to our model.
The first three features (Fbank, Log-Fbank, MFCC-26) in Fig. \ref{fig:single_features} are of shape $(246,26)$ whereas the fourth feature (MFCC-13) is of the shape $(246,13)$. The proposed model is designed to be able to handle different input dimensions.  The overall results of the experiments are shown in Table \ref{result}. In the second step, we used multiple feature combination of the four selected features (Fig. \ref{fig:single_features}) for performance analysis. In this case, individual features are vertically concatenated, which increases the height of the input (minimum 26 to maximum 65) to our model. 

\subsection{Optimization and hyperparameter tuning}
We use the cross-entropy loss from the final softmax layer output and perform optimization by stochastic gradient descent (SGD) using the Nesterov momentum method \cite{nesterov1983method}. We also use a cyclic learning rate (CLR) scheduler to change the learning rate in a triangular shape with the same height. Hyper-parameters of the scheduler, which includes minimum and maximum learning rate and step size, are calculated according to \cite{smith2017cyclical}. The model generally converges to the optimal point within 50-60 epochs while trained using these parameters. Majority voting is used to obtain the final abnormality prediction of the heart sound recordings obtained from individual 2.5s segments. 

\subsection{Performance metrics}
The final decision provided by the network is used to compute the following performance matrices: AUC, F1-score, and modified accuracy (Macc)\cite{humayun2020towards}. The Macc value 
is obtained by averaging sensitivity and specificity \cite{liu2016open}, and thus can be considered a reliable metric for the overall system performance. 

In addition, we also calculate the mean and STD of the accuracy metric for each system within individual data subsets \{a-f\} \cite{liu2016open} distributed as shown in Fig. \ref{fig:model}. These data subsets are collected in different environmental locations (varying noisy conditions) and using different stethoscopes/sensors (varying convolutional channel effect) \cite{liu2016open} and are summarized in Table \ref{data}. This subset-wise performance evaluation is essential as it shows if the overall performance is being skewed towards a specific subset due to data imbalance as seen in Fig. \ref{fig:model}. The mean and STD accuracy over the subsets measure the robustness and consistency of the performance across different domains.
\renewcommand{\arraystretch}{1.5}
\begin{table*}[tbh]
    \setlength{\tabcolsep}{6pt}
    \centering
    \caption{Experimental results of the baseline and proposed systems for the heart sound abnormality detection task.}
    \resizebox{\linewidth}{!}{
        \begin{tabular}{@{}cccccccccc@{}}
            \toprule
            \multirow{2}{*}{\textbf{Method}} & 
                \multirow{2}{*}{\textbf{AUC}} & 
                \multirow{2}{*}{\textbf{F1 Score}} & 
                \multirow{2}{*}{\textbf{Macc}} & 
                \multicolumn{5}{c}{\textbf{Accuracy in data subsets / domains}} & 
                \multirow{2}{*}{\textbf{Domain avg. acc.}} \\\cmidrule(lr){5-9} &
                & & & \textbf{a} & \textbf{b} & \textbf{c} & \textbf{d} & \textbf{e} \\\midrule 
            \multicolumn{10}{c}{\cellcolor[HTML]{EFEFEF}\textit{\textbf{\hl{1D-CNN based End-to-End systems}}}} \\
            Potes-CNN \cite{potes2016ensemble} & 79.50 & 74.99 & 73.50 & 60.00 & 63.26 & 71.43 & 40 & 100 & 66.94$\pm$21.80 \\
            Potes-CNN DBT \cite{humayun2020towards} & 84.20 & 80.67 & 79.79 & 71.25 & 70.41 & 85.71 & 66.66 & 97.5 & 78.31$\pm$12.95 \\
            Humayun \emph{tConv}-CNN DBT \cite{humayun2020towards} & 83.04 & 81.91 & 81.49 & 76.25 & 71.43 & 71.42 & 80 & 97.75 & 79.37$\pm$9.73 \\
            
            \multicolumn{10}{c}{\cellcolor[HTML]{EFEFEF}\textit{\textbf{\hl{2D-CNN (ResNet) based systems using spectro-temporal features}}}} \\
            Fbank & 89.33 & 82.18 & 82.72 & 71.25 & 76.53 & 71.43 & 90 & 100 & 81.84$\pm$11.36 \\
            Log-Fbank & 90.37 & \textbf{84.40} & 84.55 & 76.25 & \textbf{77.55} & 85.71 & 80 & 100 & 83.90$\pm$8.68 \\
            MFCC-26 & 87.82 & 79.54 & 81.15 & 77.50 & 69.39 & 85.71 & 80 & 100 & 82.52$\pm$10.20 \\
            MFCC-13 & 89.07 & 84.07 & 83.63 & 73.75 & 76.53 & 85.71 & 80 & 100 & 83.20$\pm$9.30 \\
            MFCC-13+$\Delta$  & 89.53 & 80.30 & 81.56 & 71.25 & 72.45 & 85.71 & 90 & 100 & 83.88$\pm$10.87 \\
            MFCC-13+$\Delta$+$\Delta\Delta$ & 87.51 & 78.29 & 80.09 & 72.50 & 68.37 & 85.71 & 80 & 100 & 81.32$\pm$11.10 \\
            \multicolumn{10}{c}{\cellcolor[HTML]{EFEFEF}\textit{\textbf{\hl{2D-CNN (ResNet) based systems using feature-level fusion (Proposed)}}}} \\
            Fbank \& Log-Fbank  & 90.67 & 83.22 & 83.18 & 72.50 & 75.51 & 85.71 & 90 & 100 & 84.74$\pm$9.96 \\
            Fbank \& MFCC-13  & \textbf{91.36} & 84.09 & \textbf{85.08} & \textbf{78.75} & 76.53 & \textbf{85.71} & \textbf{90} & \textbf{100} & \textbf{86.20}$\pm$8.42 \\
            Log-Fbank \& MFCC-13  & 89.93 & 82.53 & 83.37 & 73.75 & 76.53 & 85.71 & 80 & 100 & 83.20$\pm$9.30 \\
            Fbank, Log-Fbank \&  MFCC-13  & 90.61 & 82.84 & 83.71 & 77.50 & 75.51 & 85.71 & 80 & 98.88 & 83.52$\pm$8.41\\
            \bottomrule
        \end{tabular}}
        \label{result}
\end{table*}

\subsection{Baselines methods and implementation}
In this work, we use the best performing system in the Physionet 2016 CinC Challenge developed by Potes et al. \cite{potes2016ensemble}. This branched CNN model (Potes-CNN) is implemented as described in \cite{potes2016ensemble} and with the addition of the proposed DBT training scheme to address the domain variability present in the dataset. The network uses a set of finite impulse response (FIR) filters as a filterbank front-end and provides inferences for each segmented cardiac cycle \cite{humayun2020towards}. The branched CNN model \cite{potes2016ensemble} with including the DBT method to handle domain variability is denoted as the Potes-CNN-DBT method.
The results of these systems are summarized in Table \ref{result}.

\subsection{Performance of individual features}
First, we look at the performance of the individual features in the second section of Table \ref{result}. The best performing individual feature is the Log-Fbank with AUC, F1-score, and Macc values of $90.37\%$, $84.40\%$, and $84.55\%$, respectively. 
\begin{figure}[t]
\centering 
\includegraphics[width=\linewidth]{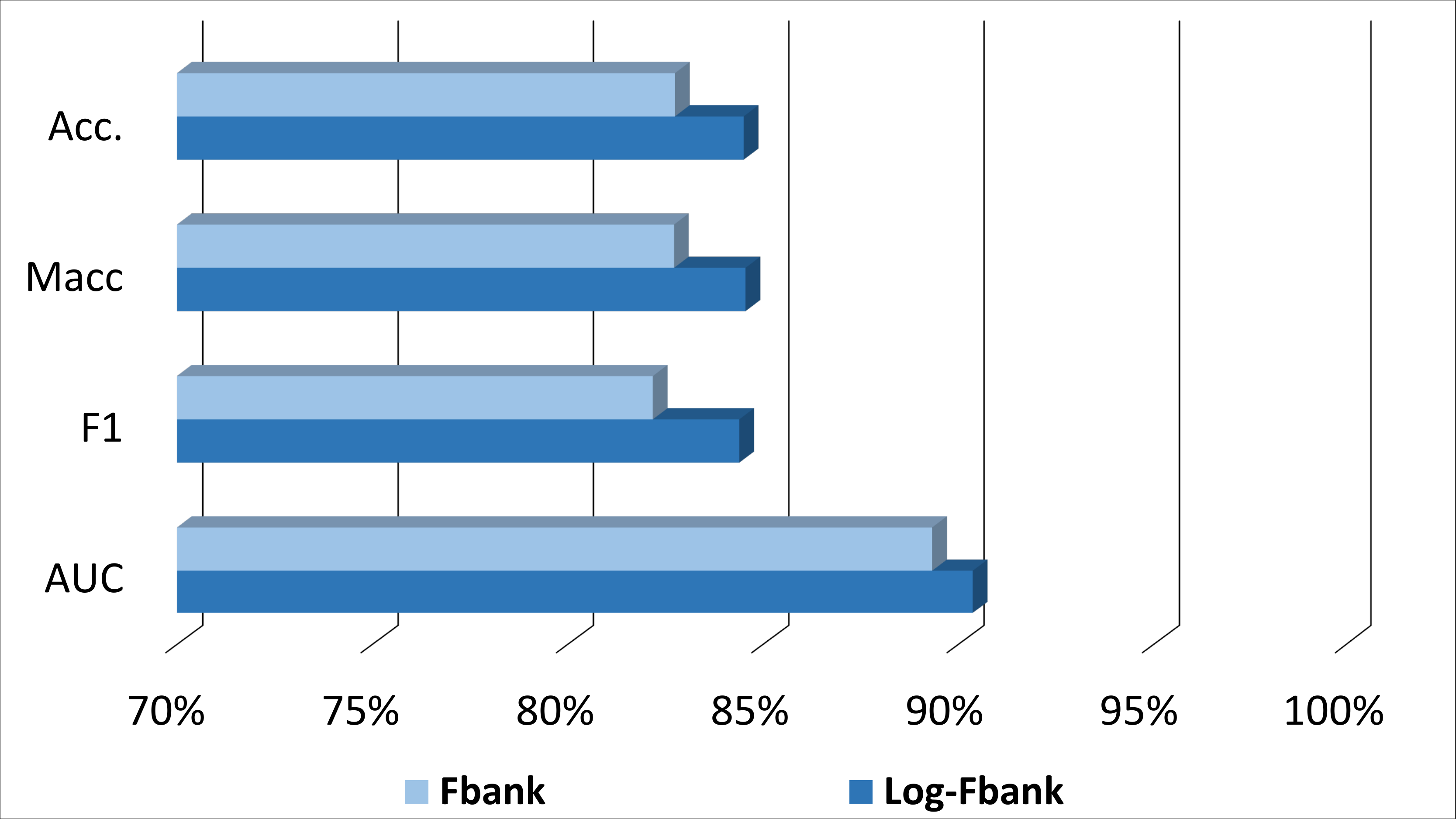}
\caption{Comparison of performance between Fbank and Log-Fbank features depicting the effect of logarithm operation on the features.}
\label{fig:log}
\end{figure}
We hypothesized the logarithm operation on the Fbank features would improve the system performance since it can express the sensor variation as an additive component in the feature space.
Observing the results of the Fbank and Log-Fbank features separately in Fig. \ref{fig:log}, we indirectly validate this hypothesis as the logarithm operation provides an overall improvement in the classification performance while also improving the robustness of the system across the data subsets (increase in average domain-wise accuracy and reducing the STD). 
It should be emphasized that both the features Fbank and Log-Fbank carry the same information content since logarithm is a one-to-one function. Therefore, we assume the improvement is due to the model's ability to fit the data better.

\begin{figure}[t]
\centering 
\includegraphics[width=\linewidth]{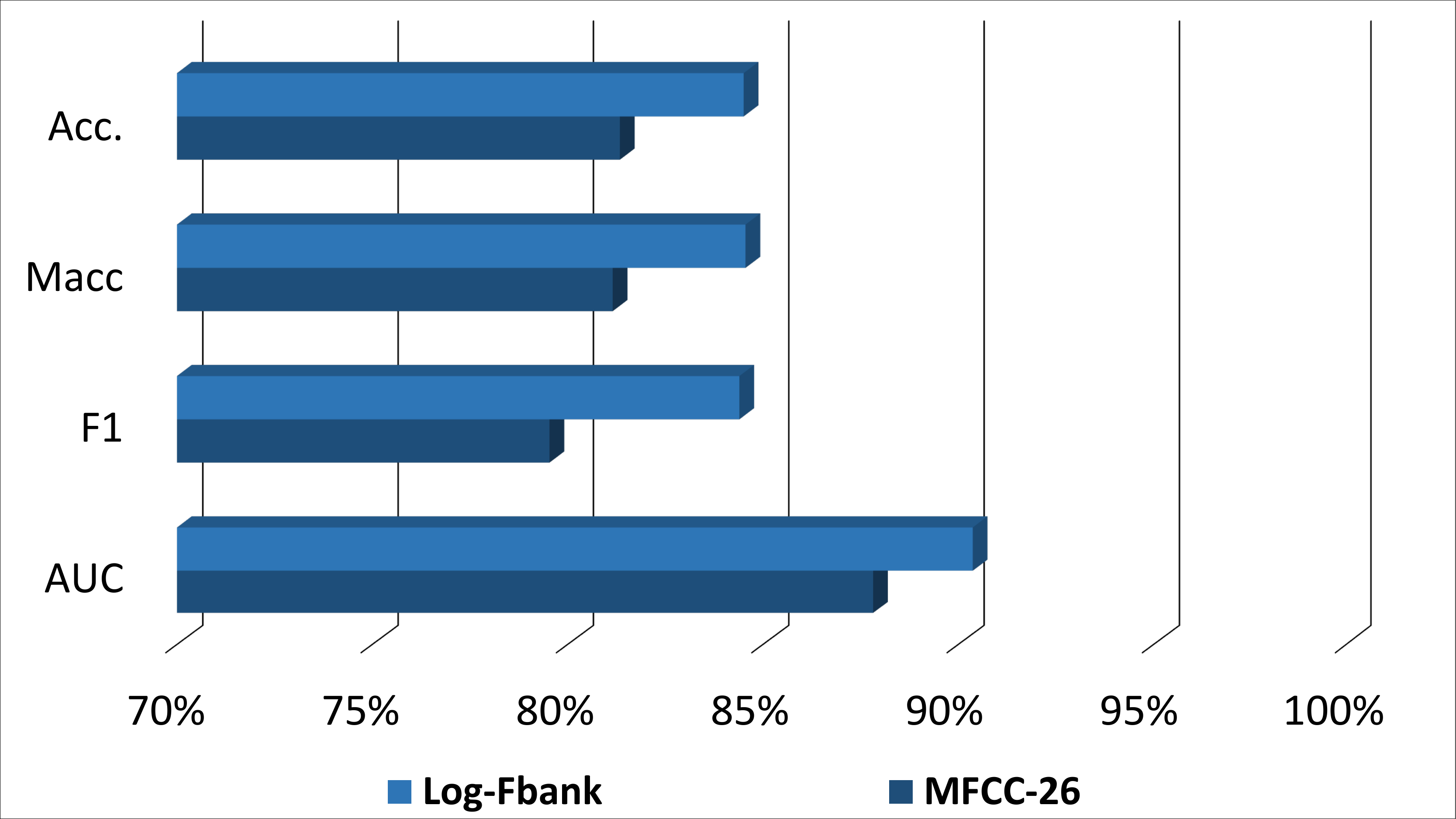}
\caption{Comparison of performance between Log-Fbank and MFCC-26 features depicting the effect of DCT operation without dimensionality reduction.}
\label{fig:dct}
\end{figure}

Next, we analyze the performance of the system after the DCT operation leading to MFCC-26, and later MFCC-13 due to dimensionality reduction. The DCT operation is generally used in feature dimensionality reduction as it performs compression of the data into a few coefficients and also de-correlates them. In theory, deep learning architectures should be able to model complex relation among the input features and thus decorrelating the features using a data-independent linear transformation, i.e. DCT, should not have an impact. However, in practice we observe that most of the performance metrics drop due to the DCT operation without dimensionality reduction. The performance of Log-Fbank and MFCC-26 are graphically compared in Fig. \ref{fig:dct} where we can better observe the performance difference.

The performance drop from Log-Fbank to MFCC-26 may be explained as follows. Since the DCT operation performs feature compression, the high frequency DCT coefficients should be of of negligible value and carry very little information about the PCG signal \cite{deller2000discrete}. Although the 26 components of Log-Fbank and MFCC-26 theoretically contain the exact same information, since the features are normalized before being fed in to the ResNet model, it is possible that these insignificant frequency components are being amplified out of proportion and confusing the model. This is consistent with our observation that by removing the 13 higher order coefficients from MFCC-26, we improve the performance as evident from the results of the MFCC-13 feature vector. This feature set provides the best overall performance among the individual features with a sensitivity, specificity, F1-score and Macc values of $89.86\%$, $77.40\%$, $84.07\%$, and $83.63\%$, respectively. 
The performance of MFCC-13 closely follows as expected since this feature configuration widely validated in speech and audio processing applications \cite{deller2000discrete, hansen2015speaker}.

\subsection{Effect of velocity and acceleration}
\begin{figure}[tbh]
\centering 
\includegraphics[width=\linewidth]{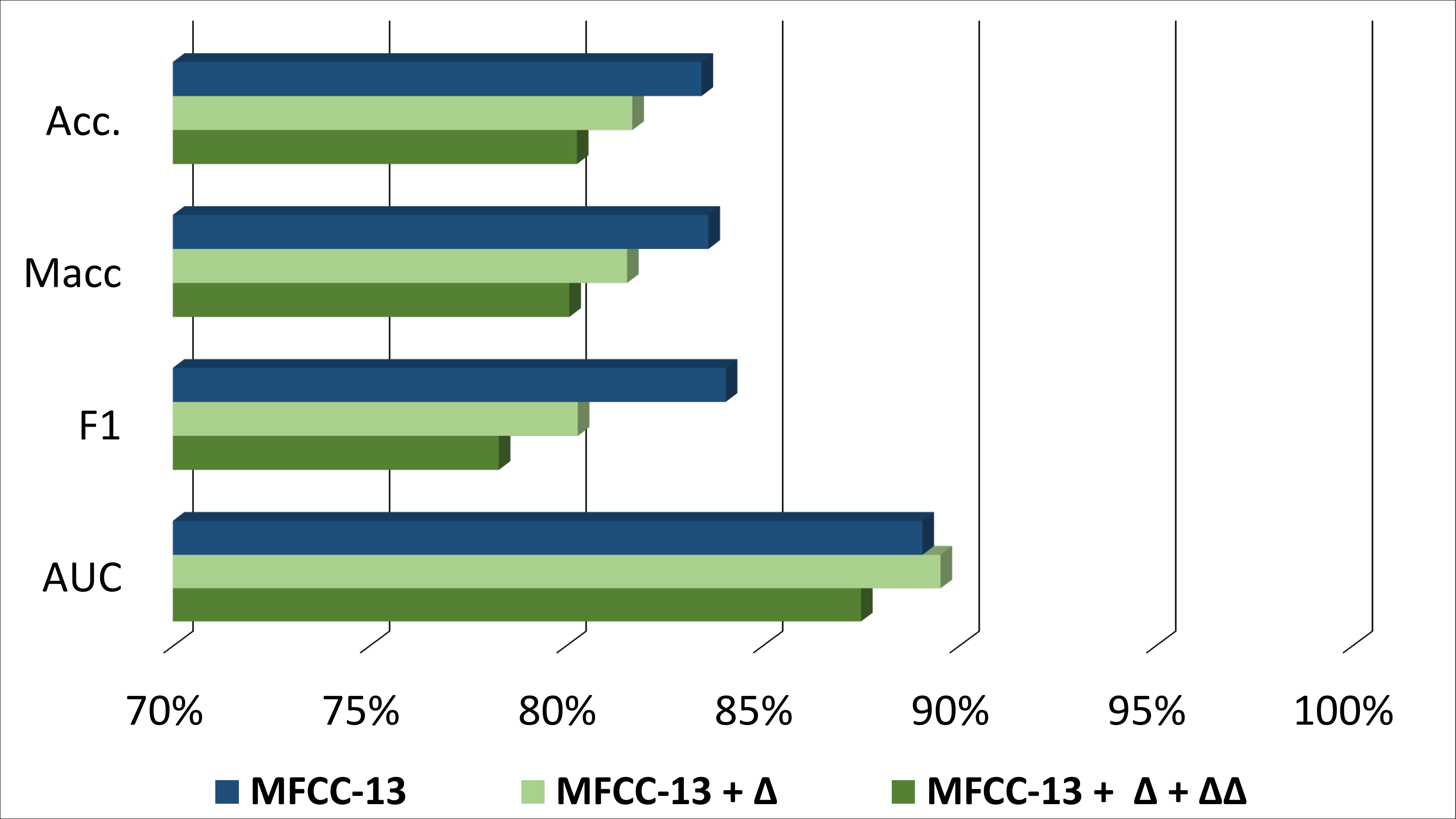}
\caption{Comparison of performance among MFCC-13, MFCC-13+$\Delta$ and $\Delta$+$\Delta\Delta$ features depicting the effect of appending velocity and acceleration features on the heart sound classification task.}
\label{fig:delta}
\end{figure}

Appending the velocity ($\Delta$) and acceleration features ($\Delta\Delta$) is very common along with MFCC-13 features, and thus we also examine the effect of these simple feature modifications. From the results in Table \ref{result} and Fig. \ref{fig:delta}, we observe that the effect of adding these features do not improve the performance of the static MFCC-13 coefficients. 
We presume that since the CNN model inputs the successive MFCC-13 vectors as a 2D image, it can model the velocity and acceleration features quite well, and thus the addition of these features is redundant.

\subsection{Feature level fusion}
The previous sub-sections analysis reveals that Log-Fbank and MFCC-13 are the best two features for heart sound abnormality detection for classification using the ResNet model \hl{described}. In Sec. \ref{motivation_fuse} we argued that Fbank and Log-Fbank features have complimentary properties since they can express additive and convolutional distortions, respectively, as an additive component in the feature space. Therefore, our hypothesis is that fusion of the Fbank and Log-Fbank features would significantly improve heart sound classification performance, particularly on this task since it involves both stethoscope/sensor and background noise variability in the dataset \cite{liu2016open}. However, in practice, the MFCC-13 and MFCC-26 features have the same property of expressing convolution as a linear operation, as shown in (\ref{eq_MFCC_additive}), and thus can be equally effective in the presence of sensor variability.

In this section of our analysis, we discuss our feature-level fusion experiments, as shown in the third section of Table \ref{result}. These experiments use several combinations of Fbank, Log-Fbank, and MFCC-13 features, being the most promising individual feature sets. The results show that, indeed, the feature-level fusion of Fbank and MFCC-13 provides the overall best results and the domain-wise performance. This configuration yields an Macc score of $85.08\%$ which shows an absolute gain of $5.29\%$ and $11.58\%$ over the Potes-CNN \cite{potes2016ensemble} and Potes-CNN DBT \cite{humayun2020towards} baseline systems, respectively. In terms of domain-wise performance, it performs better in almost all of the data subsets \{a-e\} with a mean accuracy of $86.2\% (\pm 8.42\%)$. Thus, the average accuracy over different domains has increased, but the STD of the accuracy is also \hl{reduced compared to individual spectro-temporal features}. This validates our hypothesis that this particular feature combination is complementary and effectively addresses the issues of convolutional (sensor variability) and additive distortion (noise) components.

\begin{figure}
\centering 
\includegraphics[width=\linewidth]{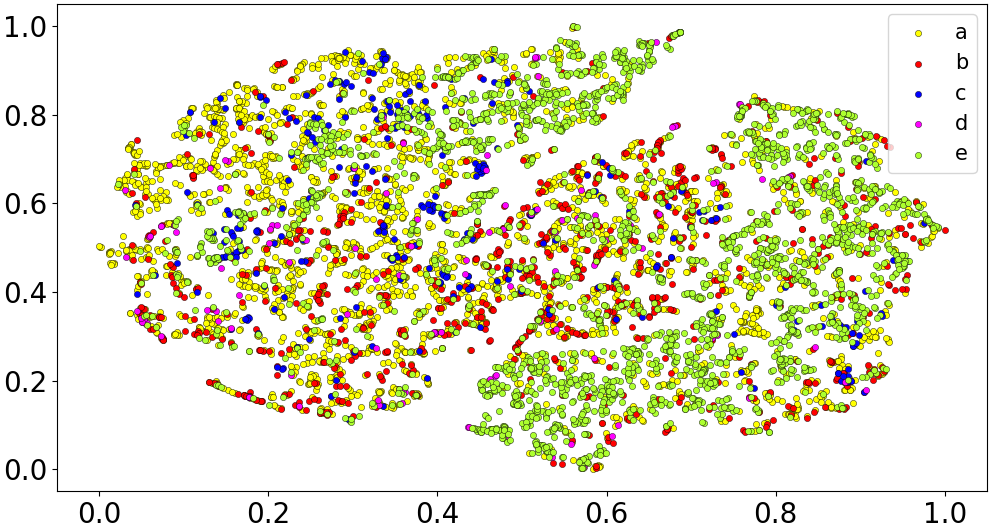}
\caption{TSNE visualization of the last layer of the proposed model showing the distribution of different data subsets. No meaningful clusters have formed with the features of different source domains.}
\label{fig:tsne}
\end{figure}

We want to point out several other observations from the results in Table \ref{result}. Firstly, the feature level fusion of all three features (Fbank, Log-Fbank, MFCC-13) does not show any additional benefit, as evident from Fig. \ref{result}. Secondly, a significant improvement in reducing domain-variability is achieved by introducing the DBT training scheme, which was already noted in \cite{humayun2020towards}. This is evident from the reduction of STD values in the domain-wise performance evaluation. The Potes-CNN \cite{potes2016ensemble} with and without DBT yields domain-wise accuracies of $66.94\% (\pm 21.80)$ and $78.31\% (\pm 12.95\%)$, respectively. Therefore, the results have already been improved by the DBT scheme. However, the proposed feature fusion scheme provides an additional improvement over the Potes-CNN DBT system, which further validates the effectiveness of the proposed method.
We have also provided the t-distributed Stochastic Neighbor Embedding (t-SNE) of the extracted features of our model trained with this fused feature set in Fig. \ref{fig:tsne}, where we observe that the different domains are almost indistinguishable in this higher-dimensional space. This depiction is consistent with the observations presented in \cite{humayun2020towards}.

            
\begin{table}[t]
    \centering
    \caption{\emph{McNemar's Chi-squared test} for statistical significance analysis compared to the baseline Potes-CNN-DBT \cite{potes2016ensemble,humayun2020towards} system.}
    \label{mcnemar}
    \resizebox{\linewidth}{!}{
    \begin{tabular}{@{}cccc@{}} 
            \toprule
            \multirow{2}{*}{\textbf{Input Feature}} &            \multirow{2}{*}{\textbf{$\chi^2$}} & 
            \multirow{2}{*}{\textbf{p-value}} & 
            \textbf{Significant}\\
                &&&($p<0.05$)\\\midrule
            Fbank & 24.00 & 0.2892 & No  \\
            Log-Fbank & 24.00 & 0.0980 & No\\
            MFCC-26 & 0.28 & 0.5962 & No \\
            MFCC-13 & 21.00 & 0.1690 & No \\
            MFCC-13+$\Delta$  & 0.4464 & 0.5040 & No\\
            MFCC-13+$\Delta$+$\Delta\Delta$ & 0.0167 & 0.8973 & No\\
            \midrule
            Fbank \& Log-Fbank  & 29.00 & 0.2750 & No\\
            Fbank \& MFCC-13  & 18.00 & \bf{0.0365} & \textbf{Yes}\\
            Log-Fbank \& MFCC-13  & 19.00 & 0.1524 & No\\
            Fbank, Log-Fbank \&  MFCC-13  & 18.00 & 0.1114 & No\\
            \bottomrule
        \end{tabular}}
\end{table}
\subsection{Statistical significance analysis}
In this section, we perform statistical significance tests to evaluate if the improvements obtained by the proposed method is significantly better than the baseline or not. We use McNemar's chi-squared test \cite{dietterich1998approximate} to perform a pair-wise comparison of the system performances and validate \hl{whether} the system performance is significantly different. The results are summarized in Table \ref{mcnemar}. According to this analysis, it is indeed confirmed that the performance improvement achieved by the proposed method with fusion of Fbank and MFCC-13 is significant ($p<0.05$) compared to the baseline system Potes-CNN DBT \cite{potes2016ensemble, humayun2020towards}. This feature-fusion combination provides the lowest p-value among all the other system comparisons and further justifies the effectiveness of the method in heart sound datasets with sensor and noise variability. The proposed system also provides an absolute $3.59\%$ improvement in Macc compared to Humayun \emph{tConv}-CNN DBT \cite{humayun2020towards}. However, this improvement is not significant according to the McNemar's chi-squared test ($p>0.05$).

\begin{figure}
\centering 
\includegraphics[width=\linewidth]{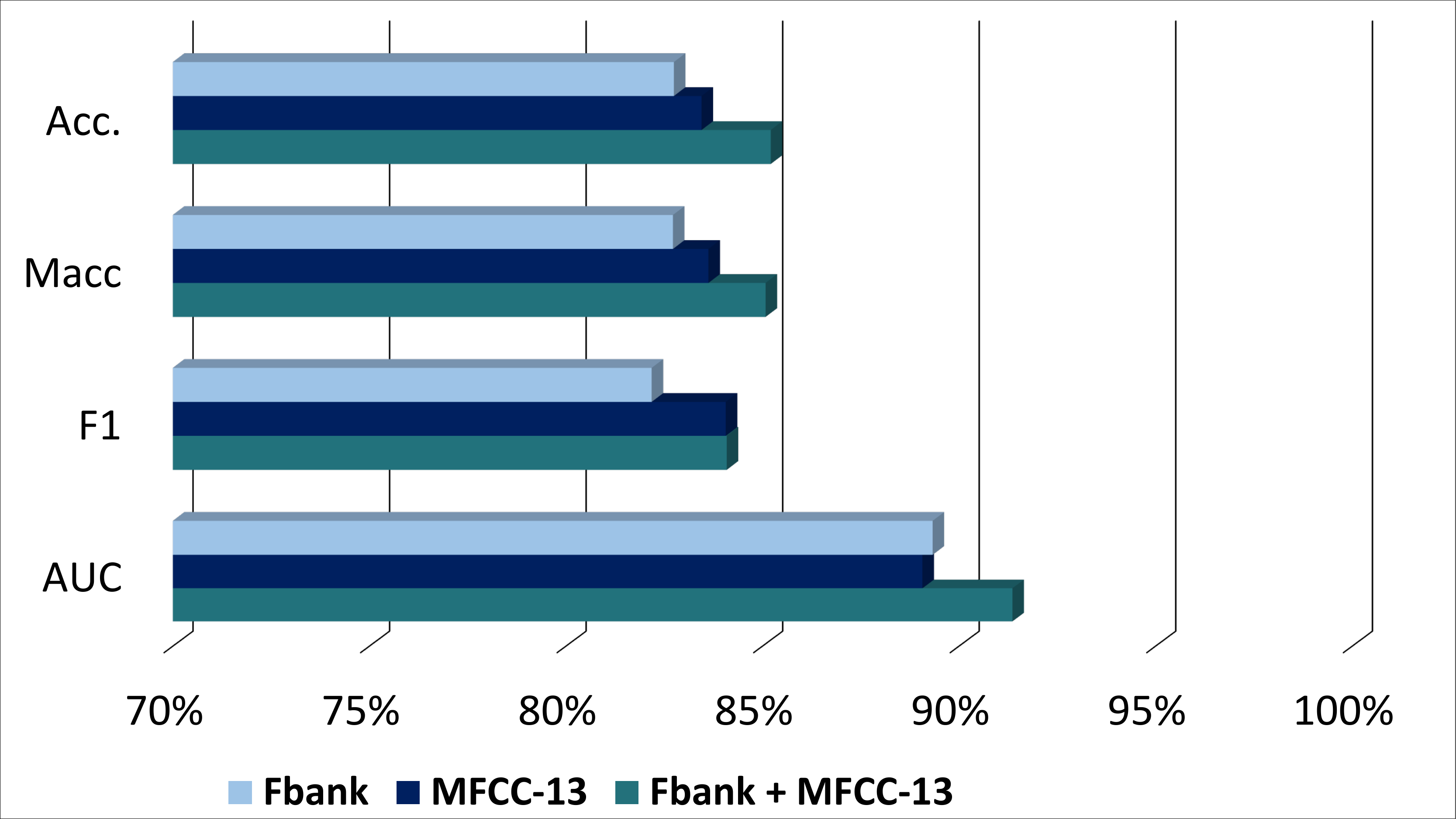}
\caption{Improvement obtained by fusing linear and logarithmic features (Fbank and MFCC-13) in heart sound abnormality detection in the presence of noise and sensor variation.}
\label{fig:merged}
\end{figure}

\section{Discussion} \label{disc}
In this work, we have indirectly validated the hypothesis that feature-level fusion of filter-bank energy features and log-filter-bank energy features provide robust performance in the heart sound abnormality detection task in the presence of noise and sensor variability. \hl{However, a major limitation of the proposed approach is that the validation is empirical by nature.} \hl{Although it is known that the Physionet data \cite{liu2016open} contains environmental noise,} it is not possible to precisely identify the sensor dependent (convolutional distortion) and the background noise (additive distortion) components from the PCG signals. However, since the results achieved are significantly improved compared to existing methods, especially methods that already deal with reducing domain-variability \cite{humayun2020towards}, the proposed method is promising and justifies further investigation. 
\hl{A more thorough analysis could have been performed by artificially generating stethoscope effects using its impulse response \cite{rennoll2020electronic} and adding environmental noise to clean heart sound recordings. In such a scenario, it would have been possible to observe the effect of the features in the presence or absense of additive and convolutional distortions more systematically. However, such analysis is outside the scope of this current work where we focus on a novel feature fusion scheme to handle channel and noise variability. The proposed method is simple and can also be applied to other biomedical signal analysis tasks that involve noise and channel distortion.}

\section{{Conclusions}} \label{conc}
This paper has first mathematically analyzed the simultaneous effect of additive and convolutional distortions for heart sound abnormality detection. In the context of PCG signals, we have hypothesized that the additive distortion components represent environmental and other body sounds, whereas convolutional distortion includes sensor variability and transmission channel effects. We have shown that effectively designed features using a combination of filter-bank energy and its natural logarithm can significantly reduce domain variability due to noise and sensor degradation in a heart sound classification task. A residual network architecture has been proposed and evaluated to classify the feature-stream cascaded to form 2D input feature matrices. Experimental results have demonstrated that the proposed method achieves a significant ($p<0.05$) absolute improvement of $5.29\%$ in Macc performance metric compared to a competitive baseline system that utilizes a domain-balanced training (DBT) scheme.
\ifCLASSOPTIONcaptionsoff
  \newpage
\fi
\bibliographystyle{IEEEtran}
\bibliography{refs.bib}

\begin{thebibliography}{10}
\providecommand{\url}[1]{#1}
\csname url@samestyle\endcsname
\providecommand{\newblock}{\relax}
\providecommand{\bibinfo}[2]{#2}
\providecommand{\BIBentrySTDinterwordspacing}{\spaceskip=0pt\relax}
\providecommand{\BIBentryALTinterwordstretchfactor}{4}
\providecommand{\BIBentryALTinterwordspacing}{\spaceskip=\fontdimen2\font plus
\BIBentryALTinterwordstretchfactor\fontdimen3\font minus
  \fontdimen4\font\relax}
\providecommand{\BIBforeignlanguage}[2]{{%
\expandafter\ifx\csname l@#1\endcsname\relax
\typeout{** WARNING: IEEEtran.bst: No hyphenation pattern has been}%
\typeout{** loaded for the language `#1'. Using the pattern for}%
\typeout{** the default language instead.}%
\else
\language=\csname l@#1\endcsname
\fi
#2}}
\providecommand{\BIBdecl}{\relax}
\BIBdecl

\bibitem{whofact}
W.~H.~O. fact~sheet 317. (2017, May) {Cardiovascular diseases (CVDs)}. [Online]
  Available:
  \url{https://www.who.int/en/news-room/fact-sheets/detail/cardiovascular-diseases-(cvds)}.
  Accessed: September 2019.

\bibitem{HeartDis39:online}
{Centers for Disease Control (CDC)}, ``Heart disease facts | cdc.gov,''
  \url{https://www.cdc.gov/heartdisease/facts.htm}, Sep 2020, (Accessed on
  09/13/2020).

\bibitem{leng2015electronic}
S.~Leng, R.~San~Tan, K.~T.~C. Chai, C.~Wang, D.~Ghista, and L.~Zhong, ``The
  electronic stethoscope,'' \emph{Biomed. Eng. Online}, vol.~14, no.~1, pp.
  1--37, 2015.

\bibitem{west2019introducing}
E.~West, I.~McLane, D.~McLane, D.~Emmanouilidou, M.~Elhilali, J.~E. West,
  A.~Ward, I.~Busch-Vishniac, J.~McLane, and B.~Dottin-Haley, ``Introducing
  feelix, a digital stethoscope incorporating active noise control and
  automatic detection of lung sound abnormalities,'' \emph{J. Acoust. Soc.
  Am.}, vol. 145, no.~3, pp. 1923--1923, 2019.

\bibitem{mclane2021design}
I.~M. Mclane, D.~Emmanouilidou, J.~West, and M.~Elhilali, ``Design and
  comparative performance of a robust lung auscultation system for noisy
  clinical settings,'' \emph{IEEE Journal of Biomedical and Health
  Informatics}, 2021.

\bibitem{goldberger2000physiobank}
A.~L. Goldberger, L.~A. Amaral, L.~Glass, J.~M. Hausdorff, P.~C. Ivanov, R.~G.
  Mark, J.~E. Mietus, G.~B. Moody, C.-K. Peng, and H.~E. Stanley, ``Physiobank,
  physiotoolkit, and physionet: components of a new research resource for
  complex physiologic signals,'' \emph{Circulation}, vol. 101, no.~23, pp.
  e215--e220, 2000.

\bibitem{schuller2017interspeech}
B.~Schuller, S.~Steidl, A.~Batliner, E.~Bergelson, J.~Krajewski, C.~Janott,
  A.~Amatuni, M.~Casillas, A.~Seidl, M.~Soderstrom \emph{et~al.}, ``The
  interspeech 2017 computational paralinguistics challenge: Addressee, cold \&
  snoring,'' in \emph{Interspeech 2017}, 2017, pp. 3442--3446.

\bibitem{clifford2016classification}
G.~D. Clifford, C.~Liu, B.~Moody, D.~Springer, I.~Silva, Q.~Li, and R.~G. Mark,
  ``Classification of normal/abnormal heart sound recordings: The
  physionet/computing in cardiology challenge 2016,'' in \emph{2016 Computing
  in cardiology conference (CinC)}.\hskip 1em plus 0.5em minus 0.4em\relax
  IEEE, 2016, pp. 609--612.

\bibitem{liu2016open}
C.~Liu, D.~Springer, Q.~Li, B.~Moody, R.~A. Juan, F.~J. Chorro, F.~Castells,
  J.~M. Roig, I.~Silva, A.~E. Johnson \emph{et~al.}, ``An open access database
  for the evaluation of heart sound algorithms,'' \emph{Physiol. Meas.},
  vol.~37, no.~12, p. 2181, 2016.

\bibitem{homsi2017ensemble}
M.~N. Homsi and P.~Warrick, ``Ensemble methods with outliers for
  phonocardiogram classification,'' \emph{Physiol. Meas.}, vol.~38, no.~8, p.
  1631, 2017.

\bibitem{rubin2016classifying}
J.~Rubin, R.~Abreu, A.~Ganguli, S.~Nelaturi, I.~Matei, and K.~Sricharan,
  ``Classifying heart sound recordings using deep convolutional neural networks
  and mel-frequency cepstral coefficients,'' in \emph{2016 Computing in
  cardiology conference (CinC)}.\hskip 1em plus 0.5em minus 0.4em\relax IEEE,
  2016, pp. 813--816.

\bibitem{bobillo2016tensor}
I.~J.~D. Bobillo, ``A tensor approach to heart sound classification,'' in
  \emph{Proc. IEEE CinC}.\hskip 1em plus 0.5em minus 0.4em\relax IEEE, 2016,
  pp. 629--632.

\bibitem{bozkurt2018study}
B.~Bozkurt, I.~Germanakis, and Y.~Stylianou, ``A study of time-frequency
  features for cnn-based automatic heart sound classification for pathology
  detection,'' \emph{Computers in biology and medicine}, vol. 100, pp.
  132--143, 2018.

\bibitem{noman2019short}
F.~Noman, C.-M. Ting, S.-H. Salleh, and H.~Ombao, ``Short-segment heart sound
  classification using an ensemble of deep convolutional neural networks,'' in
  \emph{ICASSP 2019-2019 IEEE International Conference on Acoustics, Speech and
  Signal Processing (ICASSP)}.\hskip 1em plus 0.5em minus 0.4em\relax IEEE,
  2019, pp. 1318--1322.

\bibitem{ren2018learning}
Z.~Ren, N.~Cummins, V.~Pandit, J.~Han, K.~Qian, and B.~Schuller, ``Learning
  image-based representations for heart sound classification,'' in
  \emph{Proceedings of the 2018 International Conference on Digital Health},
  2018, pp. 143--147.

\bibitem{kay2017dropconnected}
E.~Kay and A.~Agarwal, ``Dropconnected neural networks trained on
  time-frequency and inter-beat features for classifying heart sounds,''
  \emph{Physiol. Meas.}, vol.~38, no.~8, p. 1645, 2017.

\bibitem{whitaker2017combining}
B.~M. Whitaker, P.~B. Suresha, C.~Liu, G.~D. Clifford, and D.~V. Anderson,
  ``Combining sparse coding and time-domain features for heart sound
  classification,'' \emph{Physiol. Meas.}, vol.~38, no.~8, p. 1701, 2017.

\bibitem{zabihi2016heart}
M.~Zabihi, A.~B. Rad, S.~Kiranyaz, M.~Gabbouj, and A.~K. Katsaggelos, ``Heart
  sound anomaly and quality detection using ensemble of neural networks without
  segmentation,'' in \emph{Proc. IEEE CinC}.\hskip 1em plus 0.5em minus
  0.4em\relax IEEE, 2016, pp. 613--616.

\bibitem{potes2016ensemble}
C.~Potes, S.~Parvaneh, A.~Rahman, and B.~Conroy, ``Ensemble of feature-based
  and deep learning-based classifiers for detection of abnormal heart sounds,''
  in \emph{Proc. IEEE CinC}.\hskip 1em plus 0.5em minus 0.4em\relax IEEE, 2016,
  pp. 621--624.

\bibitem{humayun2020towards}
A.~I. Humayun, S.~Ghaffarzadegan, M.~I. Ansari, Z.~Feng, and T.~Hasan,
  ``Towards domain invariant heart sound abnormality detection using learnable
  filterbanks,'' \emph{IEEE J. Biomed. Health Inform.}, 2020.

\bibitem{maknickas2017recognition}
V.~Maknickas and A.~Maknickas, ``Recognition of normal--abnormal
  phonocardiographic signals using deep convolutional neural networks and
  mel-frequency spectral coefficients,'' \emph{Physiol. Meas.}, vol.~38, no.~8,
  p. 1671, 2017.

\bibitem{humayun2018learning}
A.~I. Humayun, S.~Ghaffarzadegan, Z.~Feng, and T.~Hasan, ``Learning front-end
  filter-bank parameters using convolutional neural networks for abnormal heart
  sound detection,'' in \emph{2018 40th Annual International Conference of the
  IEEE Engineering in Medicine and Biology Society (EMBC)}.\hskip 1em plus
  0.5em minus 0.4em\relax IEEE, 2018, pp. 1408--1411.

\bibitem{humayun2018ensemble}
A.~I. Humayun, M.~Khan, S.~Ghaffarzadegan, Z.~Feng, T.~Hasan \emph{et~al.},
  ``An ensemble of transfer, semi-supervised and supervised learning methods
  for pathological heart sound classification,'' \emph{arXiv preprint
  arXiv:1806.06506}, 2018.

\bibitem{yang2016classification}
T.-c.~I. Yang and H.~Hsieh, ``Classification of acoustic physiological signals
  based on deep learning neural networks with augmented features,'' in
  \emph{Proc. IEEE CinC}.\hskip 1em plus 0.5em minus 0.4em\relax IEEE, 2016,
  pp. 569--572.

\bibitem{deng2020heart}
M.~Deng, T.~Meng, J.~Cao, S.~Wang, J.~Zhang, and H.~Fan, ``Heart sound
  classification based on improved mfcc features and convolutional recurrent
  neural networks,'' \emph{Neural Networks}, vol. 130, pp. 22--32, 2020.

\bibitem{oppenheim2004frequency}
A.~V. Oppenheim and R.~W. Schafer, ``From frequency to quefrency: A history of
  the cepstrum,'' \emph{IEEE Signal Process. Mag.}, vol.~21, no.~5, pp.
  95--106, 2004.

\bibitem{acero2000hmm}
A.~Acero, L.~Deng, T.~Kristjansson, and J.~Zhang, ``Hmm adaptation using vector
  taylor series for noisy speech recognition,'' in \emph{Proc. ISCA ICSLP},
  2000.

\bibitem{stouten2004joint}
V.~Stouten, H.~Van~Hamme, and P.~Wambacq, ``Joint removal of additive and
  convolutional noise with model-based feature enhancement,'' in \emph{2004
  IEEE International Conference on Acoustics, Speech, and Signal Processing},
  vol.~1.\hskip 1em plus 0.5em minus 0.4em\relax IEEE, 2004, pp. I--949.

\bibitem{gong2005method}
Y.~Gong, ``A method of joint compensation of additive and convolutive
  distortions for speaker-independent speech recognition,'' \emph{IEEE
  transactions on speech and audio processing}, vol.~13, no.~5, pp. 975--983,
  2005.

\bibitem{deller2000discrete}
J.~R. Deller, J.~G. Proakis, and J.~H. Hansen, \emph{Discrete-time processing
  of speech signals}.\hskip 1em plus 0.5em minus 0.4em\relax IEEE, 2000.

\bibitem{liu1993efficient}
F.-H. Liu, R.~M. Stern, X.~Huang, and A.~Acero, ``Efficient cepstral
  normalization for robust speech recognition,'' in \emph{HUMAN LANGUAGE
  TECHNOLOGY: Proceedings of a Workshop Held at Plainsboro, New Jersey, March
  21-24, 1993}, 1993.

\bibitem{hyder2017acoustic}
R.~Hyder, S.~Ghaffarzadegan, Z.~Feng, J.~H. Hansen, and T.~Hasan, ``Acoustic
  scene classification using a cnn-supervector system trained with auditory and
  spectrogram image features.'' in \emph{Interspeech}, 2017, pp. 3073--3077.

\bibitem{tran2017bayesian}
T.~Tran, T.~Pham, G.~Carneiro, L.~Palmer, and I.~Reid, ``A bayesian data
  augmentation approach for learning deep models,'' in \emph{Advances in neural
  information processing systems}, 2017, pp. 2797--2806.

\bibitem{mikolajczyk2018data}
A.~Miko{\l}ajczyk and M.~Grochowski, ``Data augmentation for improving deep
  learning in image classification problem,'' in \emph{Proc. IEEE
  IIPhDW}.\hskip 1em plus 0.5em minus 0.4em\relax IEEE, 2018, pp. 117--122.

\bibitem{springer2015logistic}
D.~B. Springer, L.~Tarassenko, and G.~D. Clifford, ``Logistic
  regression-hsmm-based heart sound segmentation,'' \emph{IEEE. Trans. Biomed.
  Eng.}, vol.~63, no.~4, pp. 822--832, 2015.

\bibitem{he2016deep}
K.~He, X.~Zhang, S.~Ren, and J.~Sun, ``Deep residual learning for image
  recognition,'' in \emph{Proc. IEEE CVPR}, 2016, pp. 770--778.

\bibitem{glorot2010understanding}
X.~Glorot and Y.~Bengio, ``Understanding the difficulty of training deep
  feedforward neural networks,'' in \emph{Proc. AiSTATS}, 2010, pp. 249--256.

\bibitem{nesterov1983method}
Y.~E. Nesterov, ``A method for solving the convex programming problem with
  convergence rate o (1/k\^{} 2),'' in \emph{Dokl. akad. nauk Sssr}, vol. 269,
  1983, pp. 543--547.

\bibitem{smith2017cyclical}
L.~N. Smith, ``Cyclical learning rates for training neural networks,'' in
  \emph{Proc. IEEE WACV}.\hskip 1em plus 0.5em minus 0.4em\relax IEEE, 2017,
  pp. 464--472.

\bibitem{hansen2015speaker}
J.~H. Hansen and T.~Hasan, ``Speaker recognition by machines and humans: A
  tutorial review,'' \emph{IEEE Signal Process. Mag.}, vol.~32, no.~6, pp.
  74--99, 2015.

\bibitem{dietterich1998approximate}
T.~G. Dietterich, ``Approximate statistical tests for comparing supervised
  classification learning algorithms,'' \emph{Neural Comp.}, vol.~10, no.~7,
  pp. 1895--1923, 1998.

\bibitem{rennoll2020electronic}
V.~Rennoll, I.~M. McLane, D.~Emmanouilidou, J.~West, and M.~Elhilali,
  ``Electronic stethoscope filtering mimics the perceived sound characteristics
  of acoustic stethoscope,'' \emph{IEEE Journal of Biomedical and Health
  Informatics}, 2020.

\end{thebibliography}
\EOD
\end{document}